\documentclass[aip,jcp,reprint]{revtex4-1}
\usepackage{float}
\usepackage{graphicx}
\usepackage{grffile}
\usepackage{multirow}
\usepackage{gensymb}
\usepackage{booktabs}
\usepackage[table]{xcolor}
\usepackage{tabularx}
\usepackage{footnote}
\usepackage{amsmath} 
\usepackage{color,amsmath,amssymb}
\usepackage{color}
\usepackage{comment}
\usepackage{amssymb}
\usepackage{subcaption}
\usepackage{tikz}
\usepackage{wrapfig}

\newcommand{\bohr}{a$_{\text{o}}$}
\newcommand{\wavenumber}{cm$^{-1}$}

\begin{document}

\title{Use of the complete basis set limit for computing highly accurate \textit{ab initio} dipole moments.}

\author{Eamon K. Conway\footnote{To whom correspondence should be addressed; email: eamon.conway@cfa.harvard.edu}}
 \address{Harvard \& Smithsonian $|$ Center for Astrophysics,  Atomic and Molecular Physics Division, Cambridge, MA, USA. 02138}
\address{Department of Physics and Astronomy, University College London, Gower Street, London WC1E 6BT, United Kingdom}

\author{Iouli E. Gordon}
 \address{Harvard \& Smithsonian $|$ Center for Astrophysics,  Atomic and Molecular Physics Division, Cambridge, MA, USA. 02138}

\author{Oleg L. Polyansky}
\address{Department of Physics and Astronomy, University College London, Gower Street, London WC1E 6BT, United Kingdom}

\author{Jonathan Tennyson}
\address{Department of Physics and Astronomy, University College London, Gower Street, London WC1E 6BT, United Kingdom}

\begin{abstract}

Calculating dipole moments with high-order basis sets is generally only possible for the light molecules, such as water. A simple, yet highly effective strategy of obtaining high-order dipoles with small, computationally less expensive basis sets is described. 
Using the finite field method for computing dipoles, energies calculated with small basis sets can be extrapolated to produce dipoles that are comparable to those obtained in high order calculations. The method reduces computational resources by approximately 50\% (allowing the calculation of reliable dipole moments for larger molecules) and simultaneously improves the agreement with experimentally measured infrared transition intensities.
For atmospherically important molecules which are typically too large to consider the use of large basis sets, this procedure will provide the necessary means of improving calculated spectral intensities by several percent.

\end{abstract}

\maketitle

\section{Introduction}

The ability of atmospheric science missions to accurately detect trace amounts of molecules in our atmosphere is placing significant demands on spectroscopy \cite{tempo2,sciamanchy,omi_aura,Sentinel-4,GOME}, both theoretical and experimental. For instance, spectrometers on board of satellites including GOSAT\cite{GOSAT} and OCO-2 \cite{OCO2} now aim to detect carbon dioxide in the terrestrial atmosphere to an accuracy better than 0.3\%. Achieving this precision requires all line parameters to be known to very high accuracy. Atmospheric missions rely on the spectroscopic data in the HITRAN molecular database \cite{jt691s} which typically combines experimental and theoretical data in order to achieve both completeness and accuracy of reference line lists. Every four years the HITRAN parameters are being updated and extended to incorporate state of the art data. 

Modern experimental methods, especially those that employ frequency combs (see for instance \cite{VASILCHENKO2019332,MIKHAILENKO2018163,MONDELAIN2017206}) can measure line positions with errors on the order of 10$^{-5}$ \wavenumber, if not smaller, which is several orders of magnitude better than the accuracy that can be achieved by calculations using the best semi-empirical potential energy surfaces available \cite{12HuScTa.CO2,11BuPoZo.H2O,jt714}. For example, for water vapor the most accurate potential energy surface available \cite{jt714} predicts energy levels below 15~000 \wavenumber\ with a standard deviation $\sigma = 0.011$ \wavenumber. These high-accuracy semi-empirical potentials are achieved via a process known as refinement \cite{03YuCaJe.PH3}. When experimental data is available, the parameters used in fitting the underlying electronic structure calculations are adjusted to improve the agreement between theory and the observed data. This procedure, when done correctly, allows energy levels to be predicted to less than 0.03 \wavenumber, hence the accuracy is no longer proportional to the level of theory considered for the electronic structure calculations \cite{11BuPoZo.H2O,jt561,jt714, jt734}. However, when refinement is not possible, then the level of theory becomes significantly more important. Extrapolating calculated energies to the complete basis set (CBS) is one such method that has been used with great success. Second order corrections such as relativistic, adiabatic, non-adiabatic, quantum electrodynamics and spin orbit coupling can also become important \cite{schwenkenbo,jt309,jt274,jt550}.

Cavity ring down spectroscopy can measure those transitions that have very weak intensities with high accuracy \cite{MIKHAILENKO2019106574}. For well studied molecules including H$_{2}$O and CO$_{2}$, transition intensities obtained from \textit{ab initio} calculations and experiments are generally in excellent agreement with deviation on the sub one-percent scale for many bands \cite{jt613,jt625,jt687}.


For a given method, the basis set chosen for the electronic structure calculations largely determines the accuracy of an \textit{ab initio} dipole calculation.  The Dunning \cite{89Dunning.ai,95WoDuxx.ai,02PeDuxx.ai} aug-cc-pCV6Z basis set is the largest conventional basis set yet considered for the calculation of a global dipole surface \cite{jt744,jt509} for a triatomic system, in this case water. Each of these dipole calculations required almost two days worth of CPU time to compute \cite{jt509, jt744}, hence for molecules with more electrons, such as CO$_2$, use the aug-cc-pCV6Z basis is currently unfeasible.

Even with the extensive computer resources used in the creation of accurate dipole moment surfaces (DMS) for water, over hundred years of CPU time in the case the CKAPTEN DMS,\cite{jt744} 
there still remains issues with predicted infrared transition intensities. For example, the $\nu_{1}$ fundamental has been shown to be very sensitive to the choice of \textit{ab initio} calculations \cite{97PaScxx.H2O,00ScPaxx.H2O, jt744, jt687} and improvements are needed. Computing dipoles with a larger basis set is one solution, although not feasible on a global scale. Similar problems persist for other atmospherically important molecules such as CO$_2$ \cite{19HuScLe.CO2} and ozone \cite{jt721,19TyBaJa.O3}.

This work aims explores the use of complete basis set (CBS) extrapolation in the computation of  high-accuracy \textit{ab initio} spectral intensities both to give increased accuracy and to reduce the computational cost. 

\section{Method}

\subsection{Calculations}

We have computed over 16000 aug-cc-pCV6Z \cite{89Dunning.ai,95WoDuxx.ai,02PeDuxx.ai} finite-field dipoles for water using the electronic structure program MOLPRO\cite{MOLPRO}. With these we created a DMS for water vapor, named CKAPTEN \cite{jt744}. Spectra computed with this CKAPTEN surface have been shown to produce excellent transition intensities when compared to both experiment and observation \cite{jt775}. From these 16000 configurations, we create a smaller sub-grid with 2540 data points. The points were chosen such that there is sufficient coverage of both stretching and bending coordinates, which would provide accurate transition intensities up to approximately 15000 cm$^{-1}$. For calculating reliable spectra in the ultraviolet, it is necessary to limit the use of fitting parameters as done in CKAPTEN, which reduces the possibility of artificial oscillations occurring in highly energetic regions of the dipole surface. In this work, we are not interested in such energetic transitions and are not required to use a few parameter fit nor to use 16000 points to better constrain the fit. The stretching coordinates in this sub-grid are in the range of 1.4 \bohr\ to 4 \bohr, while the angular selection lies between 30$^{\text{o}}$ and 178$^{\text{o}}$. With respect to our equilibrium configuration of 1.8141 \bohr\ and 104.52$^{\text{o}}$, our data set includes dipoles with energies up to 42481 \wavenumber. The original 16000 finite-field dipoles were computed at the multi-reference configuration interaction (MRCI) level the aug-cc-pCV6Z  basis set, a Davidson correction (+Q), Douglass-Kroll-Hess-Hamiltonian to order two (DKH2), and with an electric field strength of 5$\times$10$^{-5}$ a.u. We calculate finite-field dipoles on our sub-grid of 2540 points that use smaller basis sets, notably the aug-cc-pCV(Q,5)Z sets, while still using the same formalism employed for the aug-cc-pCV6Z dipoles. 
Calculations were all performed on Legion, Grace and Myriad systems at the University College London High Performance Computing Facilities and CPU times given below are for these computers.

\subsection{Extrapolation Technique}

We configure our water molecule such that the z-axis bisects the angle HOH, with the x-axis in the HOH plane and  perpendicular to z. In the finite field approach, we apply a small electric field, $\lambda$, in each of the directions: $+\hat{x}$, $-\hat{x}$, $+\hat{z}$ and $-\hat{z}$. The energy of the molecule in each respective field will be termed: $E_{px}$,  $E_{nx}$,  $E_{pz}$ and $E_{nz}$. $p$ and $n$ signify positive and negatively directed fields, while $x$ and $z$ are the components.  Each dipole component is calculated as: 
\begin{equation}
    \mu_{x(z)} = \frac{(E_{px(pz)} - E_{nx(nz)})}{2\lambda}
\end{equation} 
We extrapolate the individual energies $E_{px}$,  $E_{nx}$,  $E_{pz}$ and $E_{nz}$ with a standard expression \cite{cbs_water}:

\begin{equation}
E_{x}=E_{cbs}+be^{-x}    
\end{equation}
where $E_{cbs}$ and $b$ are fitted parameters while $x$ represents the level of theory. This will provide us with $E^{cbs}_{px}$,  $E^{cbs}_{nx}$,  $E^{cbs}_{pz}$ and $E^{cbs}_{nz}$. This formula (2) has successfully been used in the past to extrapolate \textit{ab initio} energies for potential energy surfaces \cite{jt394,11BuPoZo.H2O,jt550}. Instead of one single energy extrapolation to do, we have five: the zero-field calculation and one for each of our four dipole components within an electric field.

We are interested in applying this technique to produce extrapolated dipoles, $ \mu^{cbs_{Q5}}_{x(z)}$ and $ \mu^{cbs_{56}}_{x(z)}$. We expect the $ \mu^{cbs_{Q5}}_{x(z)}$ dipoles to behave as the aug-cc-pCV6Z dipoles, and for the $ \mu^{cbs_{56}}_{x(z)}$ dipoles to behave as aug-cc-pCV7Z dipoles.

To create $ \mu^{cbs_{Q5}}_{x(z)}$ dipoles, we require $E^{cbs_{Q5}}_{px}$,  $E^{cbs_{Q5}}_{nx}$,  $E^{cbs_{Q5}}_{pz}$ and $E^{cbs_{Q5}}_{nz}$, which represent the individually extrapolated energies. Likewise, to calculate $ \mu^{cbs_{56}}_{x(z)}$, we need $E^{cbs_{56}}_{px}$,  $E^{cbs_{56}}_{nx}$,  $E^{cbs_{56}}_{pz}$ and $E^{cbs_{56}}_{nz}$. Combining these values, we can now calculate our extrapolated dipoles: 
\begin{equation}
    \mu^{cbs}_{x(z)} = \frac{(E^{cbs}_{px(pz)} - E^{cbs}_{nx(nz)})}{2\lambda}
\end{equation}

In total, we possess five sets of dipoles, one for each of the aug-cc-pCV(Q,5,6)Z basis sets and then two sets of extrapolated sets, CBS$^{\text{Q}5}$ and CBS$^{56}$.

\subsection{Fitting}

We fit each dipole set to a similar functional form used in the creation of CKAPTEN\cite{jt744}:

\begin{equation}
 	\mu_{z}(r_{1},r_{2},\theta)=(\pi - \theta)\sum\limits_{{i,j,k}}C_{ijk}^{(z)}\; \zeta _{1}^{i} \;\zeta_{2}^{j} \;\zeta_{3}^{k}   
\end{equation}
\begin{equation}
 	\mu_{x}(r_{1},r_{2},\theta)=\sum\limits_{{i,j,k}} C_{ijk}^{(x)}
\;\zeta_{1}^{i} \;\zeta_{2}^{j} \;\zeta_{3}^{k} 
\end{equation}
where: $\zeta_{1}=\frac{(r_{1}+r_{2})}{2}-r_{e}$,
$\zeta_{2}=(r_{2}-r_{1})$ and $\zeta_{3}=\theta / \theta_{e}$. To physically model the dipole surface correctly, we have several conditions that must be adhered to: \newline (i) $\mu_{z}(r_{1},r_{2},\theta=\pi) = 0$ \newline (ii) $\mu_{z}(r_{1},r_{2},\theta) = \mu_{z}(r_{2},r_{1},\theta)$  \newline (iii) $\mu_{x}(r_{1},r_{2},\theta) = -\mu_{x}(r_{2},r_{1},\theta)$. \newline
These requirements mean that only even (odd) powers of $j$ are used in the expression $\zeta^{j}$ for the $\hat{z}$ ($\hat{x}$) components.

 \begin{figure*}[ht]
	\includegraphics[width=1.0\linewidth]{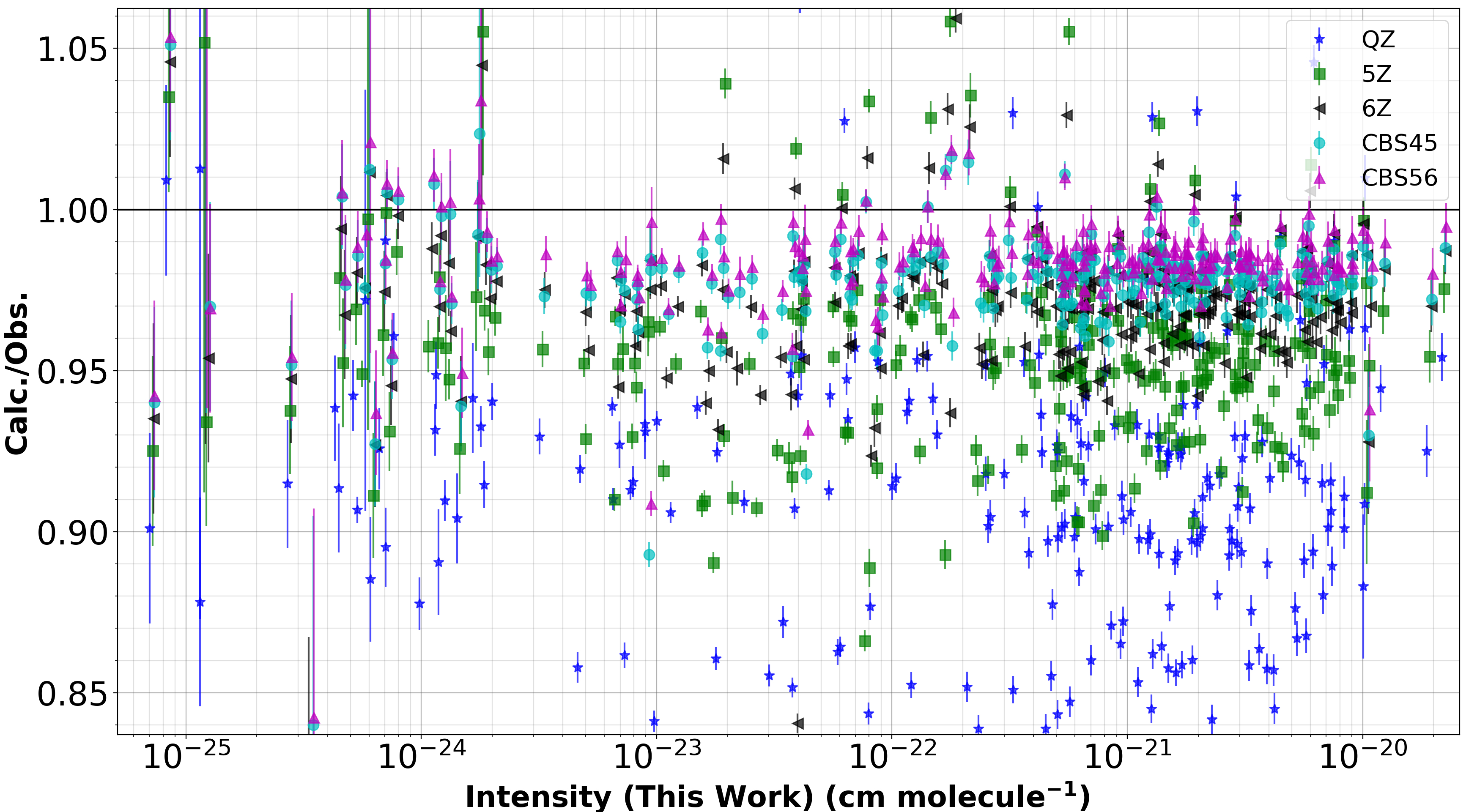}
	 \caption{ Ratios of calculated $\nu_{1}$ transition intensities obtained from each dipole moment surface against  experimental measurements of Birk \textit{et al.} \cite{16BiLoWa.H2O} and Loos \textit{et al.}  \cite{17LoBiWa.H2O} The error bars are experimental. }
	\label{fig:1}
\end{figure*}

The dipoles in each set are weighted \cite{97PaScxx.H2O,00ScPaxx.H2O} ($w_{i}$) as a function of their energy: 

\begin{multline}
s_{i} =  \text{tanh}(-0.006(E_{i}-30000)+1.002002002 )/2.002002002\\
w_{i} = 25000(s_{i})/\text{max}(E_{i},25000)
\end{multline}
To facilitate an equal comparison between the different levels of theory, each dipole set must be fit to the same functional form, requiring the same number of parameters in each fit. We fit the $\hat{z}$ components with 228 parameters and the $\hat{x}$ with 163 parameters. 

Table \ref{tab:table1} shows the average root-mean-square deviation of the dipole fitting procedure for each of the individual surfaces. For the parallel $\hat{z}$ component, each surface is fit to an RMS under 10$^{-4}$ a.u, while for the perpendicular $\hat{x}$ component, each surface possesses an RMS of approximately 1.1$\times 10^{-4}$ a.u. The perpendicular component is often the most difficult to fit and carries a larger a RMS than the respective parallel component of the dipole \cite{jt509,jt744}

\begin{table}[ht]
  \begin{center}
    \caption{The weighted root-mean-square (RMS) deviation of each fitted dipole component for each surface, see text
    for details. }
    \label{tab:table1}
    \begin{tabular}{m{2cm}p{2.5cm}p{2.5cm}}

	\hline\hline
	  	&  $\mu_{z}$  (a.u)  &  $\mu_{x}$  (a.u) \\
	\hline 

Q-Zeta	&$7.29\times 10^{-5}$ & $1.10\times 10^{-4}$	\\
5-Zeta 	& $5.55\times 10^{-5}$ & $9.95\times 10^{-5}$ \\
CBS(45) &$8.59\times 10^{-5}$ & $1.20\times 10^{-4}$ \\
6-Zeta	&$5.35\times 10^{-5}$ & $9.96\times 10^{-5}$ \\
CBS(56)	&$7.77\times 10^{-5}$ &$1.10\times 10^{-4}$ \\
	\hline 	\hline	 

	\end{tabular}
	\end{center}
\end{table}

To compare the surfaces, we need to assess the resulting transition intensities. To do this, we require wave-functions, which we calculate from the potential energy surface of Mizus \textit{et al.} \cite{jt714} This PES, called PES15K, is valid for energies that fall below 15000 cm$^{-1}$, which is sufficient for our study. For states falling below this threshold, PES15K predicts energies to a RMS of only 0.011 \wavenumber. Using the DVR3D \cite{jt338} suite of programs we calculate spectra for each of the five dipole surfaces with $J_{max}=6$, $\nu_{max}=15000$ cm$^{-1}$, $S_{\text{if}} \geq 10^{-30}$ cm molecule$^{-1}$ and $T=296$K.

\section{Results}

 \begin{figure*}[ht]
	\includegraphics[width=1.0\linewidth]{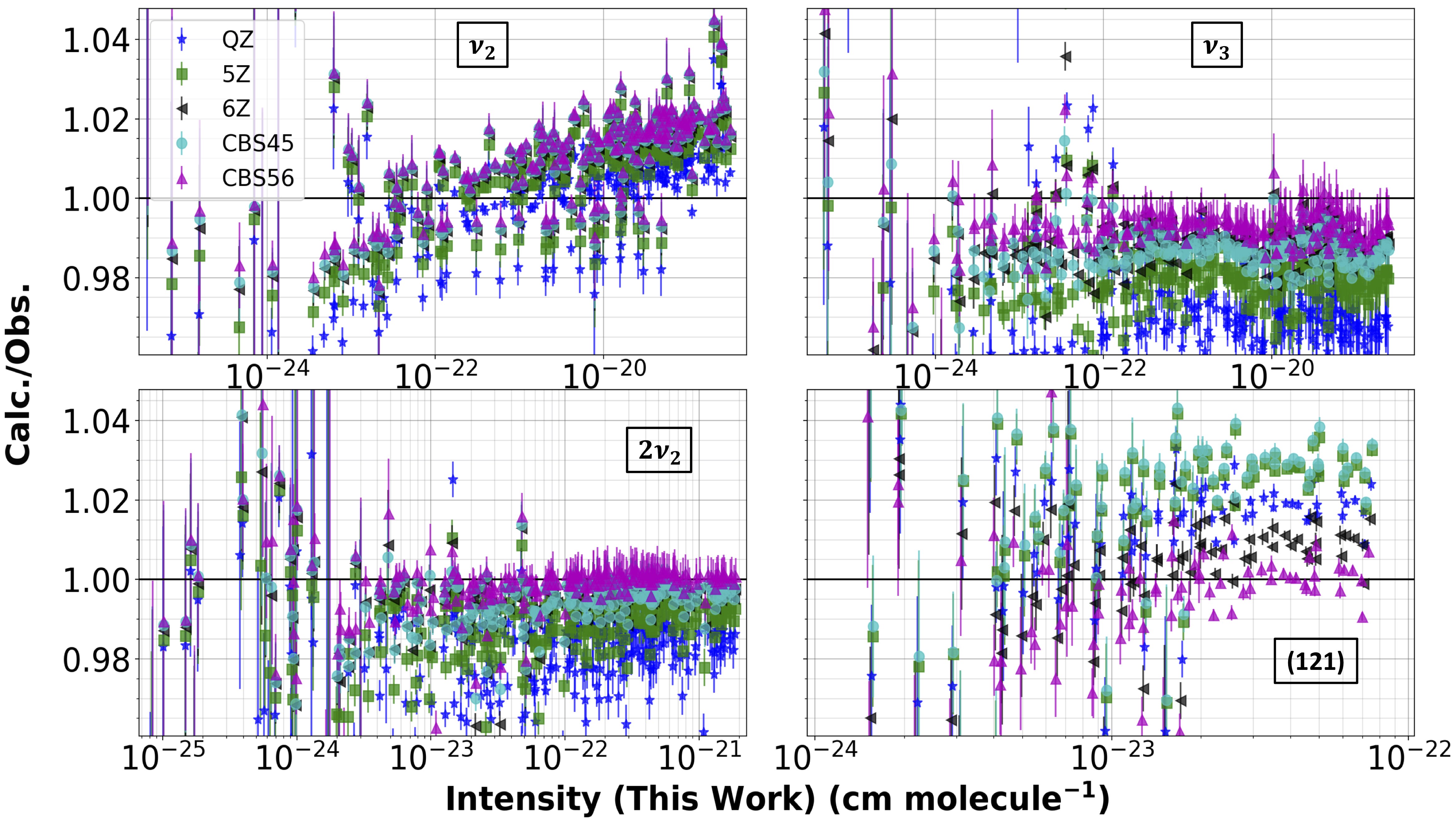}
	\caption{Ratios of intensities calculated using DMS from Table I to the experimental measurements of Birk \textit{et al.} \cite{16BiLoWa.H2O,16BiWaxx.H2O,jt687} and Loos \textit{et al.} \cite{17LoBiWa.H2O}, shown for $\nu_{3}$ \cite{16BiLoWa.H2O,17LoBiWa.H2O}, $\nu_{2}$ \cite{16BiWaxx.H2O}, 2$\nu_{2}$ \cite{17LoBiWa.H2O} and (121) \cite{jt687}}
	\label{fig:2}
\end{figure*}

Table \ref{tab:table2} contains energies $E_{px}$,  $E_{nx}$,  $E_{pz}$ and $E_{nz}$ with the corresponding dipoles, $\mu_{x}$ and $\mu_{z}$ obtained using basis sets aug-cc-pCV(X=Q,5,6)Z on one molecular geometry: $R_{1}=1.7$ \bohr, $R_{2}=1.98$ \bohr, $\theta=163^{\text{o}}$. Attempts to extrapolate dipoles $\mu_{z,x}$ for the aug-cc-pCV(X=Q,5)Z basis sets with formula $\mu_{x,z}=\mu^{cbs}_{x,z}+be^{-x}$ result in failure. The best fit to these dipoles is a linear function with $\chi^{2}$ on the order of $10^{-12}$, which is incorrect: energies can be extrapolated to their complete basis set limit, dipoles growing linearly with growth in basis set size cannot be extrapolated. 

Many extrapolation schemes exist and it is often suggested to extrapolate the Hartree-Fock energy component separately from the correlation contribution as they can converge at different rates \cite{97HeKlKo.H2O}. However, for the data in Table \ref{tab:table2}, the rate of convergence of their combined energy is described best by an exponential curve. Extrapolating the components independently with different schemes should not provide drastically different results on the dipoles, hence we estimate the fit uncertainty on the transition intensities to be on the sub-percent scale.

For the energy extrapolations, the fitted $b$ variables are found to be equal for extrapolating $E_{pz}$ and $E_{nx}$ energies, and also for $E_{px}$ and $E_{nz}$. The $\mu_{z,x}$ dipoles for QZ, 5Z and 6Z shown in Table \ref{tab:table2} were computed within the program MOLPRO using energies that were converged to the tenth significant figure, hence these dipoles are accurate to the sixth significant figure (dividing by 2$\lambda$ is equivalent to multiplying by 10000).The individual energies, $E_{px}$,  $E_{nx}$,  $E_{pz}$ and $E_{nz}$, were written out to only eight significant figures, hence the extrapolated dipoles that use these energies are instead accurate to the fourth significant figure.

Table \ref{tab:table3} shows eight selected molecular configurations and presents the dipole values calculated for each of the five theoretical methods with the average CPU time required per configuration. For the extrapolated surfaces, we combine the CPU time for each of the individual calculations required for the extrapolation to provide a total time. The CBS$^{\text{Q5}}$ dipoles deviate by less than $ 10^{-4}$ a.u from the CBS$^{\text{56}}$, which was not expected, yet these CBS$^{\text{Q5}}$ dipoles only require less than 50\% of the CPU time to compute than the aug-cc-CV6Z calculations. The marginal difference between the CBS$^{\text{Q5}}$ and CBS$^{\text{56}}$ dipoles highlights the correlation between these basis sets. This marginal deviation also holds true for those dipoles computed for geometries with a large proportion of stretch and/or bend, see Table \ref{tab:table3}, which means the technique holds true for all regions of the DMS. This result implies that highly accurate, global dipole surfaces can now be calculated with lower levels of theory at a fraction of the CPU cost, with marginal loss of precision. This should prove to be important for  molecules which are too large to compute large grids of energies with an aug-cc-pCV6Z basis set. 

\begin{table*}[ht]
  \begin{center}
    \caption{Calculated dipole moment and four energies $E_{px}$,  $E_{nx}$,  $E_{pz}$ and $E_{nz}$ with the molecular configuration consisting of $R_{1}=1.7$ \bohr, $R_{2}=1.98$ \bohr\ and $\theta=163^{\text{o}}$ computed with basis sets aug-cc-pCV(X=Q,5,6)Z. The calculated dipoles carry more precision than the energies shown here. QZ and 5Z energies are extrapolated with $E=E_{cbs}+be^{-x}$ and the respective CBS$^{\text{Q5}}$ dipole calculated.   }
    \label{tab:table2}
    \begin{tabular}{lccccc}
	\hline\hline
	  	&  QZ &  5Z  & 6Z & CBS$^{\text{Q5}}$ & $b$\\
	\hline 
\hline
$E_{pz}$		&-76.41695707   &-76.42504517  &-76.42813805 &-76.42975226   &0.69859347	\\
$E_{nz}$ &-76.41697909	&-76.42506722   &-76.42816013   &-76.42977432   &0.69859606 \\
\hline
$\mu_{z}$	&0.220221   &0.220487  & 0.220746 &0.220675 &-	\\
\hline \hline

$E_{px}$		&-76.41697514   &-76.42506327  &-76.42815617 &-76.42977037   &0.69859606	\\
$E_{nx}$		&-76.41696102   &-76.42504912  &-76.42814201  &-76.42975620  &0.69859347\\
\hline
$\mu_{x}$	&-0.141258   &-0.141482  &-0.141600 &-0.141675	&-\\
	\hline 	\hline	 

	\end{tabular}
	\end{center}
\end{table*}

In Table \ref{tab:table3}, the CBS$^{56}$ dipoles are consistently larger than those calculated with the aug-cc-pCV6Z basis set and should be of a comparable magnitude to those calculations expected with a aug-cc-pCV7Z basis set, if the smooth exponential growth of the dipole remains consistent. Obtaining these CBS$^{56}$ dipoles will be computationally expensive for all molecules, as calculations will be required at both the aug-cc-pCV5Z and aug-cc-pCV6Z levels of theory. This will be limited to the lighter molecules.

Birk \textit{et al.} \cite{jt687} analyzed transition intensities of H$_{2}$O in the IR \cite{16BiWaxx.H2O,16BiLoWa.H2O,17LoBiWa.H2O} and found the most recent (at the time) \textit{ab initio} line list to deviate from the experimental observations \cite{17LoBiWa.H2O} in the $\nu_{1}$ fundamental by 3-15\%. This \textit{ab initio} line list was calculated with the LTP2011S DMS of Lodi \textit{et al.}\cite{jt509}  Despite this large residual, all other bands in the 3000-4400 \wavenumber\ region showed good agreement with experiment, suggesting the presence of an issue in the theoretical model. We recently calculated a new line list with the CKAPTEN DMS and there were a large number of improvements over the recent POKAZATEL line list\cite{jt734}, particularly for those transitions with wavelengths below 1~$\mu$m ($\omega >$ 10000 \wavenumber).  The same conclusion was made for the isotopologues. The POKAZATEL line list was computed with a variation of the LTP2011 DMS dipole surface from Lodi \textit{et al.}, termed LTP2011S, which utilized fewer parameters for improved stability in highly energetic regions. However, our calculated intensities in the $\nu_{1}$ fundamental showed no signs of improvement over POKAZATEL, nor what was computed in 2011. Over 16000 aug-cc-pCV6Z dipoles underlay this CKAPTEN surface, hence discrepancies in this band are not due to the number of points fit, nor is it a fitting issue as different functional forms were used for each of these models. These line lists used rotation-vibration wavefunctions which were also calculated with a variety of potential energy surfaces, thus these are also not the source of the deviation, although any energy dependence in the residuals will be due to the potential.

Figure \ref{fig:1} plots the ratios of our calculated transition intensities in $\nu_{1}$ to those measurements from Loos \textit{et al.}\cite{17LoBiWa.H2O} As the level of theory increases from QZ through to CBS$^{56}$, the deviation in the $\nu_{1}$ fundamental reduces from 2.83\% with the 6Z DMS, to only 1.48\% with the new CBS$^{\text{56}}$ DMS, see Table \ref{tab:table3}. Indeed, the spectral intensities calculated with the CBS$^{\text{Q5}}$ DMS  are closer to the experimental values of Loos \textit{et al.} \cite{17LoBiWa.H2O} than the aug-cc-pCV6Z calculations. This suggests better results can potentially be obtained by extrapolating the computationally less expensive calculations. To verify the extrapolation technique works on a global scale, we need to also investigate other infrared bands measured by Loos \textit{et al.} and Birk \textit{et al.} \cite{16BiWaxx.H2O,16BiLoWa.H2O} Ratios of transition intensities in $\nu_{3}$, $\nu_{2}$ and 2$\nu_{2}$ are presented in Figure \ref{fig:2}. Also, Table \ref{tab:table3} presents a general overview of several other bands which we do not present in our figures. 

\begin{table*}[t]
		\caption{Upper: Water dipoles, in a.u., for  \textit{ab initio} calculations with different basis sets / basis set extrapolations for a range of molecular configurations. Energies ($E$) are given
		relative to the equilibrium geometry, point 1. The average CPU time required to calculate a single point is also provided. Lower:  average weighted calculated to measured intensity ratios for selected vibrational bands. 
		The experimental data are due to Birk \textit{et al.} \cite{16BiLoWa.H2O,16BiWaxx.H2O,jt687} and Loos \textit{et al.} \cite{17LoBiWa.H2O} }  
		\label{tab:table3}
		
  \begin{tabular}{lllrrccccc}
    \hline\hline
    \textbf{\#}& $\mathbf{R}_{1}$ & $\mathbf{R}_{2}$ & $\mathbf{\theta}$ & $\mathbf{E}$ \textbf{(\wavenumber)} & \textbf{QZ}  & \textbf{5Z} & \textbf{CBS}$^{\mathbf{\text{Q5}}}$ & \textbf{6Z} & \textbf{CBS}$^{\mathbf{\text{56}}}$ \\
    \hline
  1  & 1.8141    & 1.8141    & 104.52    &0    & 0.7267     & 0.7282    & 0.7291    &0.7287     &0.7290     \\
  2  & 1.68    &1.98     & 74.00    & 6190.43    & 0.8870    &  0.8890   &0.8902     & 0.8899    &0.8904     \\
  3  & 1.94    &1.94     & 59.00    & 13166.24    & 0.9432    &  0.9454   &0.9467 & 0.9462    &0.9467     \\
  4  & 2.10    &2.18 & 166.00    & 22472.55    & 0.1793    &0.1798     & 0.1801    &  0.1800   &  0.1801   \\
  5  & 2.10    & 3.00    & 105.00    &28438.03     & 0.6864    & 0.6893    & 0.6910    & 0.6904    &0.6910     \\
  6  &  1.60   &3.50     &95.00     & 37444.38    & 0.6768    &0.6792     &  0.6806   & 0.6802    & 0.6808    \\
  7  & 2.00    &4.00     & 80.00    & 39816.63    & 0.6690    & 0.6691    & 0.6704    & 0.6701    & 0.6707    \\
  8  & 2.80    &1.85     & 30.00    & 41035.46    & 0.9900    & 0.9924    &0.9938     &0.9934     &0.9940     \\
      \hline
\multicolumn{5}{c}{\textbf{CPU TIME (s)}} & 15190& 27684 &42874 & 84715 & 112399  \\
\hline
\hline
\multicolumn{5}{c}{($\mathbf{\nu_{1},\nu_{2},\nu_{3}}$)$\mathbf{'}$-($\mathbf{\nu_{1},\nu_{2},\nu_{3}}$)$\mathbf{''}$ } & &  &&&   \\
\hline
\multicolumn{5}{c}{(100)-(000) } & 8.42 &4.74  &1.99  &2.83  & 1.48  \\
\multicolumn{5}{c}{(010)-(000) } & 0.67 & 0.95 &1.16  &1.11  & 1.23  \\
\multicolumn{5}{c}{(001)-(000) } &2.44  &1.53  &0.98  & 0.89 & 0.52  \\
\multicolumn{5}{c}{(020)-(000) } & 2.17 &1.32  &0.70  & 0.73 &0.44   \\
\multicolumn{5}{c}{(121)-(000) } & 1.25 & 1.71 &1.78  & 0.80 & 0.62  \\
\hline\hline
  \end{tabular}
\end{table*}

\begin{figure}[ht]
	\includegraphics[width=1.0\linewidth]{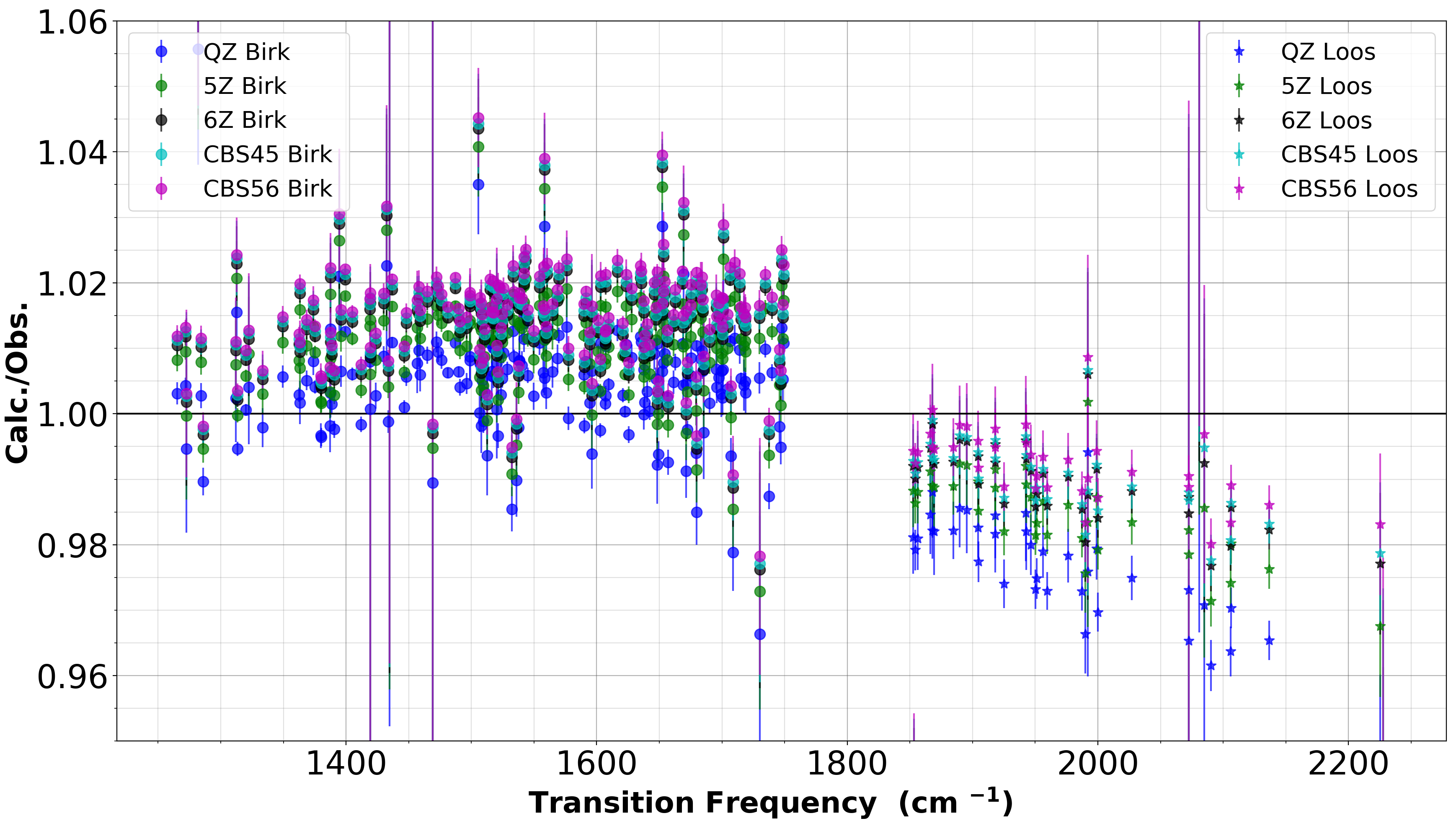}
	\caption{Ratios of intensities calculated using the DMS from Table I to the experimental measurements of Birk \textit{et al.} \cite{16BiLoWa.H2O} and Loos \textit{et al.} \cite{17LoBiWa.H2O} for $\nu_{2}$ transitions plotted as a function of frequency.}
	\label{fig:3}
\end{figure}

For the $\nu_{3}$ and 2$\nu_{2}$ bands, shown in Figure \ref{fig:2}, the computed CBS$^{\text{56}}$ spectrum is again closer to the experimental intensities than any of the other data sets. However, the $\nu_{2}$ ratios suggest the analysis of the experimental spectrum of Loos \textit{et al.} \cite{17LoBiWa.H2O} differs from that of Birk \textit{et al.} \cite{16BiLoWa.H2O}.  In Figure \ref{fig:3}, we plot intensity ratios for $\nu_{2}$ transitions as a function of frequency and separate the data sets. Birk analyzed the spectrum with a Speed-Dependent Voigt profile, while Loos used a  quadratic Speed Dependent Hard Collision profile with Rosenkranz line-mixing. Our results show that the choice of profile is very important for high-accuracy intensity measurements.

For more energetic transitions, wavelengths shorter than 1~$\mu$m, the CBS$^{56}$ intensities show excellent agreement with the measurements of Birk \textit{et al.} \cite{jt687}, see Figure \ref{fig:2}. For the same band, the calculated aug-cc-pCVQZ spectrum is closer to the experimental data than the CBS$^{\text{Q5}}$ and aug-cc-pCV5Z spectra, which is counter-intuitive. We also expect the CBS$^{\text{Q5}}$ spectrum to be closer to experiment than the aug-cc-pCV5Z spectrum, which is also not the case. The results indicate there may be an underlying issue regarding the aug-cc-pCVQZ data set in this band.

\section{Conclusion}

We compute \textit{ab initio} dipoles with the MRCI procedure embedded within the quantum chemistry package MOLPRO \cite{MOLPRO} with the aug-cc-pCV(X=Q,5,6)Z basis sets \cite{89Dunning.ai,95WoDuxx.ai,02PeDuxx.ai,MOLPRO} for water vapor. Using these results, we similarly created extrapolated CBS$^{\text{Q5}}$ and CBS$^{\text{56}}$ dipoles, each obtained by extrapolating energies from the aug-cc-pCVQZ and aug-cc-pCV5Z calculations, and likewise from the aug-cc-pCV5Z and aug-cc-pCV6Z levels of theory. The same functional form was fit to these dipole sets to create five individual dipole moment surfaces.

For a select number of global geometries studied, the extrapolated CBS$^{\text{Q5}}$ and CBS$^{\text{56}}$ dipoles are within $10^{-4}$ a.u of each other. This finding does reflect the correlation in the aug-cc-pCV(X=Q,5,6) basis sets. The greatest achievement surrounds the computational time required for calculating the CBS$^{\text{Q5}}$ dipoles: they require 50\% less CPU time compared to the aug-cc-pCV6Z dipoles. 

Using the potential energy surface of Mizus \textit{et al.}, known as PES15K, which predicts energy levels below 15000 \wavenumber\ to a root mean square of 0.011 \wavenumber, we create wave-functions up to total quantum number $J=6$. With these, we calculate spectra for each of the dipole surfaces with an upper threshold of 15000 \wavenumber\ and where possible, we replace calculated energy levels with those in the MARVEL database.  

The transition intensities obtained from each data set are compared against the experimental measurements of Birk \textit{et al.} and Loos \textit{et al.} which are both in the IR. Overall, the CBS$^{\text{56}}$ DMS, despite being fit with dipoles that are only accurate to the fourth significant figure, resulting intensities exhibit the best agreement with the measurements, and for several bands the deviation is in the sub-one-percent range. Also, comparing the CBS$^{\text{56}}$ and aug-cc-pCV6Z results indicates that it may be preferable, in cases, to consider using the CBS$^{\text{56}}$ dipoles over the aug-cc-pCV6Z dipoles. 

For those strong transitions in the IR, calculated intensities obtained using the CBS$^{\text{Q5}}$ and CBS$^{\text{56}}$ DMS's  suggests that the limit of dipole precision required for obtaining sub-percent accuracy against high-quality experiments could be on the order of 10$^{-4}$ a.u. 
	
We show the two point formula works very well for extrapolating dipoles, reducing the deviation from experimental measurements by approximately one percent in certain bands, and thus we estimate that the application of a three-point formula for dipole extrapolation could potentially improve the accuracy by another 0.5\%.

Coupled-cluster methods are in general computationally less expensive than MRCI but nonetheless provide accurate results for energies and hence finite field dipoles for geometries close to equilibrium \cite{18ChYaTe,19CoYuKo, 18OwYaTh}. The extrapolation technique is successful not because of the underlying method of calculation, in this case MRCI, but because of the nature of the aug-cc-pCV(X=Q,5,6)Z basis sets. We therefore expect the extrapolation technique to work for coupled-cluster calculations, providing the calculations remain converged.  Applying this extrapolation technique to high-level coupled-cluster energies may prove to be a cheaper alternative of obtaining mid/lower-level MRCI results.

\section*{Acknowledgments}
This was supported by the UK's Natural Environment Research Council (NERC) under grant NE/T000767/1.
The authors acknowledge the use of the UCL Grace, Legion and Myriad High Performance and High Throughput Computing Facilities (Grace@UCL, Legion@UCL, Myriad@UCL), and associated support services, in the completion of this work.

\bibliographystyle{apsrev}

\end{document}